# COLD TEST RESULTS OF PRE-PRODUCTION PIP-II SSR2 CAVITIES WITH HIGH-POWER COUPLERS IN THE FERMILAB SPOKE TEST CRYOSTAT*

A. Sukhanov†, C. Contreras-Martinez, C. Grimm, B. Hanna, B. Hansen, T. Khabiboulline, M. Parise, D. Passarelli, Y. Pischalnikov, D. Porwisiak, V. Roger, J. Subedi, A. Syed, P. Varghese, S. Wijethunga, V. Yakovlev, Fermilab, Batavia, USA

*Abstract*

As part of the PIP-II project at Fermilab, a pre-production cryomodule featuring 325 MHz Single Spoke Resonator type 2 (SSR2) superconducting RF cavities is under construction. These SSR2 cavities are fabricated by industry partners and undergo initial cold testing at our collaborating institution, IJCLab in France, utilizing low-power coupler. Subsequently, the cavities are subjected to final qualification at Fermilab, complete with tuner and high-power coupler assemblies. This paper provides an overview of the ongoing efforts dedicated to high-power testing of jacketed SSR2 cavities in the Spoke Test Cryostat (STC) at Fermilab. Performance parameters obtained from these tests are presented, offering valuable insights into the cavities' operational characteristics and readiness for integration into the PIP-II cryomodule.

## INTRODUCTION

The Proton Improvement Plan-II project (PIP-II) is built at Fermilab to deliver intense neutrino beam to LBNF-DUNE experiment [1]. PIP-II utilizes continuous-wave (CW) Superconducting RF (SRF) linac to accelerate $H^-$ ions up to 800 MeV. For efficient acceleration PIP-II linac is composed of five different types of cavities. 35 Single Spoke Resonator Type-2 (SSR2) cavities operating at 325 MHz and assembled in seven cryo-modules accelerate beam from 35 to 185 MeV [2]. Seven pre-production jacketed SSR2 cavities were fabricated by the Fermilab industry partners. Six of these cavities from one of the vendors were initially cold tested at IJCLab in France with low-power coupler (LPC). Results of this testing were discussed elsewhere [3]. SSR2 cavities from IJCLab and the one from the second vendor were delivered to Fermilab for assembly with tuner and high-power coupler (HPC) and final testing and qualification at the Fermilab Spoke Test Cryostat (STC) before installation into the pre-production SSR2 cryo-module.

The Fermilab STC facility (see Fig. 1) is capable of cold testing of four out of five types of PIP-II SRF cavities: both 325 MHz SSR type 1 [4] and type 2, and 650 MHz 5-cell elliptical Low and High beta (LB650, HB650) cavities [5]. RF tests at STC could be performed in LPC and HPC configurations. Initially, SSR2 cavities received at Fermilab from IJCLab, were tested with LPC to verify and cross-check testing procedures between different test set-ups in our institutions. Five such tests were performed on two cavities [3]. One SSR2 pre-production cavity was delivered to Fermilab recently from another vendor. Initial performance evaluation of this cavity was performed at STC in LPC setup.

After initial verification, cavities were assembled with high-power coupler and tuner. In this paper we report on ongoing efforts of HPC testing of SSR2 cavities at STC.

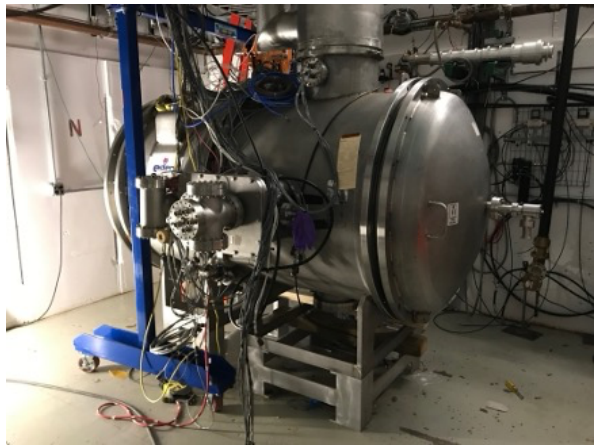

Figure 1: Spoke Test Cryostat facility at Fermilab

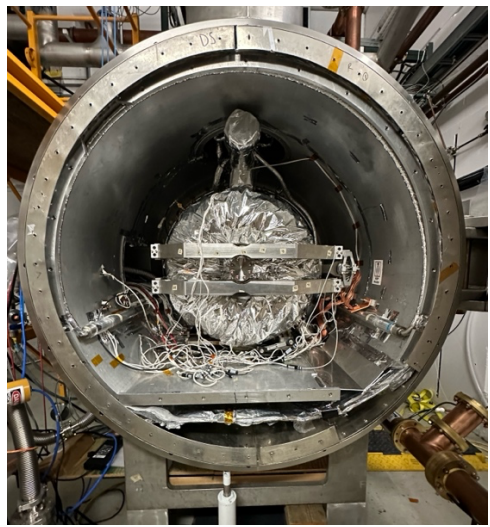

Figure 2: SSR2 Cavity with tuner and high-power coupler installed inside STC cryostat.

---

## STC TEST SEQUENCE

### Cavity Preparation and Installation

Jacketed SSR2 cavities were shipped to Fermilab assembled with LPC and under vacuum in RF/beamline volume. When no verification with LPC at STC were required, or after the STC LPC testing, cavities underwent installation of high-power coupler [6]. Cavities assembled with HPC were delivered to STC for installation of tuner and instrumentation. Typical instrumentation setup included temperature sensors at the cavity Helium vessel (top, bottom, middle), tuner arm and motor windings, coupler ceramic window flange and 5 K and 80 K intercepts; magnetic sensors at the vessel and beam pipe; heater at the coupler ceramic window flange. Subsequently, cavities were inserted into the STC cryostat, Fig. 2, where air (warm) side of the coupler was assembled and connected to the RF distribution system and thermal connection of the coupler 80 K intercept to the STC high-temperature thermal shield (HTTS) was done. Note, that because STC does not have low-temperature thermal shield, coupler 5 K intercept was not connected. After making cryogenic and beamline vacuum connections, STC cryostat was closed and insulating vacuum space was evacuated in preparation for the STC cooldown.

### Test Sequence

High-power coupler tests of SSR2 cavities consisted of the following steps: cooldown, tuner evaluation at 2 K, LLRF system calibration, coupler conditioning, cavity multipactor conditioning and characterization of cavity performance, including maximum cavity accelerating gradient, $Q_0$, radiation and limiting factor (field emission, quenches), if presented.

Typical cooldown of SSR2 cavities in STC from room temperature to 2 K took 13—15 hours. It began with cooling HTTS with $LN_2$ and Helium circuit with cold *He* gas for 10—12 hours to temperatures 120—150 K, followed by a quick 20—25 minutes transition of cooling cavity down to 5 K. Finally, Helium circuit was evacuated to 23 Torr, with cavity reaching 2 K.

Tuner evaluation and measurements were usually performed after cooldown [7], cavity resonance frequency tuned to 325 MHz and cavity loaded quality factor, $Q_L$, measured with VNA. We measured $Q_L$ = 4.3—4.5 x $10^7$ in our tests in agreement with SSR2 cavity and coupler specifications.

At the beginning of RF testing calibration of LLRF system was performed. This calibration included measurement of total attenuation in signal circuits for forward, reflected and transmitted power signals and adjustment of appropriate constats in the LLRF software. Typical total attenuation was 85 dB for forward and reflected power and 38 dB for transmitted power signals.

Coupler checkout and conditioning with cavity off resonance was performed in pulsed mode at 10 Hz with pulse length set to 2 ms, 5 ms, 10 ms, 20 ms and finally to CW mode, by gradually increasing forward power at each pulse settings up to 7 kW. During all HPC RF testing of SSR2 cavities at STC, with 4—4.5 kV HV bias always applied, no multipactoring electron activity or arching was observed in the coupler.

After coupler conditioning, cavity was tuned on resonance and cavity multipactor (MP) processing performed. With lessons learned from SSR1 prototype cavities, SSR2 cavity design [2] was optimized to shift cavity MP barriers to lower cavity field, away from the cavity operating gradient, 11.4 MV/m. We observed few relatively soft MP barriers in the ranges 0.8—1 MV/m, 1.5—1.8 MV/m and 2.7—3.3 MV/m, in agreement with design [2] and recent simulation [8]. MP conditioning was done using both pulsed and CW RF operation, with typical processing time less than 2 hours.

## RESULTS

After MP conditioning, cavity performance was evaluated. So far, two SSR2 cavities (denoted as "cavity 1" and "cavity 2" below) were tested at STC with high-power coupler. Results of these tests described below.

### Accelerating Gradient

Both cavities achieved operating gradient of 11.4 MV/m and maximum gradient of 13.7 MV/m (administrative limit). Both cavities had field emission (FE) with onset at 9 MV/m (cavity 1) and 10 MV/m (cavity 2). Radiation was measured at 2 m from cavity center along cavity axis at both ends of STC, with results shown in Fig. 3. Radiation at operating gradient was 0.7 mSv and 0.07 mSv (cavity 1 and 2, respectively). These satisfied PIP-II specifications for FE radiation (< 1 mSv).

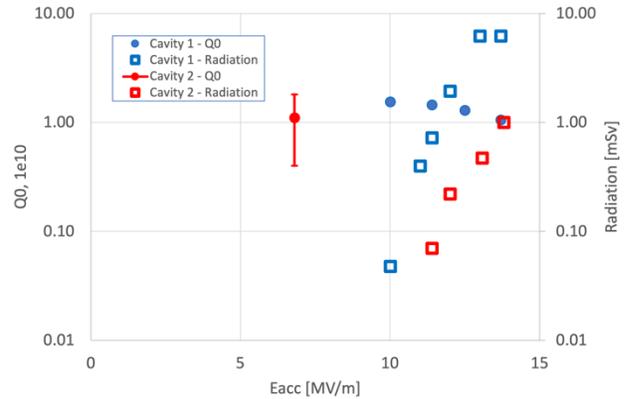

Figure 3: Intrinsic quality factor, Q0, and FE radiation for cavity 1 (blue markers) and cavity 2 (red markers)

### Cavity Quality Factor

In STC HPC testing, cavity' intrinsic quality factor, $Q_0$, was measured using "calorimetric" method, based on evaluation of the STC cryogenic system response to heat load from power loss in cavity walls. Few cryogenic parameters could be used for that purpose. In our case we relied on liquid Helium mass flow. A heater, inserted inside Helium circuit close to the cavity, was utilized for calibration of mass flow response to applied heat load. Figure 4 shows typical results that were obtained in such calibrations. Calibration constant and static mass flow were $S_M = 0.033$ (g/s)/W and $M_0 = 1.19$ g/s, respectively.

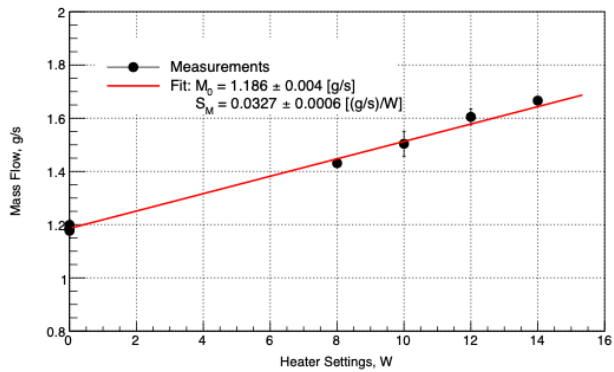

Figure 4: Heater calibration

Calorimetric method of $Q_0$ measurement was heavily dependent on performance and stability of the STC cryogenic system. Fluctuations in mass flow required longer periods of data taking (45—90 minutes) for each gradient point, with systematic uncertainty in estimation of the heat load drastically increased at lower cavity fields. Results of $Q_0$ measurements shown in Fig. 3.

Cavity 1 was measured to have $Q_0 = 1.5 \times 10^{10}$ at operating field 11.4 MV/m, satisfying PIP-II specifications (>0.9 $\times 10^{10}$). During test of cavity 2, a valve at STC cooldown line, which should be closed in normal 2 K operations, was leaking, leading to extremely large instabilities in cryogenic system behavior and unreliable measurements in heat load. Only a rough estimation of Q0 at 6.8 MV/m was obtained 1.1 $\times 10^{10}$ with a relative uncertainty of 70%. Since gradient and radiation performance of this cavity was found to be comparable with these measured during LPC test, this cavity was conditionally qualified for cryo-module integration. At the conclusion of the test, cooldown line valve was replaced and stable performance of the STC cryogenic system was later verified in the LPC test of a different SSR2 cavity.

*Lorentz Force Detuning and df/dp*

Electromagnetic field inside SRF cavity acts on the cavity walls with a force proportional to the square of the field strength, causing deformation and detuning of cavity from the nominal frequency. This called Lorentz force detuning (LFD). Variations in liquid Helium pressure inside cavity vessel may also result in detuning of the cavity resonance frequency.

Both pressure sensitivity of cavity frequency, df/dp, and LFD were measured during HPC testing of SSR2 cavities. Results are shown in Table 1.

Table 1: Pressure sensitivity and LFD factor.

|  | df/dp, Hz/mbar | LFD factor Hz/(MV/m)$^2$ |
|---|---|---|
| Cavity 1 | -3.2 | -7.4 |
| Cavity 2 | -3.6 | -7.6 |

## CONCLUSION

We reported on status of ongoing efforts of cold testing of pre-production SSR2 for PIP-II linac in STC. Both cavities reached operating and maximum gradients, 11.4 MV/m and 13.7 MV/m and satisfied PIP-II specification requirements for radiation, Lorentz Force detuning factor and pressure sensitivity of the resonance frequency. Q0 of one cavity exceeded PIP-II specification. Measurement of Q0 of the second cavity was affected by unstable performance of the STC cryogenic system. Both cavities were qualified (2nd cavity only conditionally) for the SSR2 pre-production cryomodule integration. STC cryogenic system instability problem was investigated, resolved and stable performance was verified.

We looking forward to complete HPC testing, characterization and qualification of the rest of the SSR2 cavities in next few months.